\documentclass
[12pt]
{article}
\usepackage{latexsym,amsfonts}

\makeatletter
\@addtoreset{equation}{section}
\makeatother

\def\be{\begin{equation}}
\def\ee{\end{equation}}
\def\bea{\begin{eqnarray}}
\def\eea{\end{eqnarray}}
\def\nn{\nonumber}
 
\def\c{\chi}

\def\e{\epsilon}           

\def\f{\phi}               

\def\l{\lambda}
\def\m{\mu}
\def\n{\nu}

\def\p{\pi}                
\def\r{\rho}                      
\def\s{\sigma}                    

\def\Q{\Theta}

\begin{document}

\title{\bf {A Lie-algebra model for a noncommutative space time geometry }}

\author{
B.-D. D\"orfel\\
Institut f\"ur Physik, Humboldt-Universit\"at zu Berlin\\
Invalidenstra\ss e 110, D-10115 Berlin, Germany\\}

\date{May 16, 2002}

\maketitle

\begin{abstract}

 We propose a Lie-algebra model for noncommutative coordinate and
 momentum space . Based on a rigid commutation relation for the
 commutators of space time operators the model is quite constrained if
 one tries to keep Lorentz invariance as much as possible. We discuss
 the question of invariants esp. the definition of a mass.
  
  \vskip 0.5cm

  \noindent
  Keyword(s): noncommutative geometry, Lie algebra
\end{abstract}

\section{Introduction}

After the motivation for a noncommutative geometry of space time has
been demonstrated in numerous papers (s. e. g. [1-3]) there exists now a
well established form of quantum field theory on noncommutative spaces
[4-10] which allows to calculate Feynman diagrams and related
values. Those theories show a characteristic mixing of infrared and
ultraviolet divergencies [10,11], but the study of planar diagrams exhibits
the fact that they cannot be handled without the usual ultraviolet
cutoff in momentum space necesarry for the treatment of ordinary
renormalizable quantum field theories. On the other hand it was one of
the motivations of noncommutative geometry to generate such a cutoff
within the theory what has now definitely failed for the class of
models mentioned above [11]. Therefore it seems still necessary to look for
other possibilities to employ noncommutative space time geometry. 

Generally speaking, such models are mainly contained in three classes
depending on the structure of the commutator for space time
coordinates.

a) The canonical structure:
\be [X^\l,X^\m]=i\Q^{\l\m} \ee

b) The Lie-algebra structure:
\be [X^\l,X^\m]=ic^{\l\m}_\n X^\n \ee

c) The quantum space structure:
\be [X^\l,X^\m]=id^{\l\m}_{\n\s}X^\n X^\s \ee

The last class has been intensively treated in papers like [12-15], as a
general feature space time becomes discrete avoiding most
renormalization problems for possible quantum field theories. The case
a) is usually connected with string theory while case b) has been
worked out in papers like [16-19].

Our approach shall be viewed as a combination of a) and b) because our
main ingredient is the postulate of a Lie algebra including all
operators of physical interest. As a consequence we demand the $\Q$ to
belong to that algebra as central operators. We understand that this
postulate is in disagreement with nearly all other models existing
but their drawbacks seem to provide enough arguments for testing new
conceptions. The current paper is therefore mainly devoted to
illustrating the consequences of our approach. 

Because standard quantum field theory on noncommutative spaces of type
a) is also not Lorentz covariant, it is worthwhile to re-examine the
role of Lorentz invariance in noncommutative field theory. After we
have taken as a basis the existence of a Lie algebra we found it
straightforward to demand the Lorentz group to be the second factor in
the semi-direct product of Lie groups (s. next sect.). This condition
seems neither to weak nor to strong and its consequences are worth of
being worked out.

The paper is organized as follows. In sect.2 we define our model and
determine the possible structure of $\Q$ . In sect.3 we present an
explicit representation of our noncommutative Lie algebra. Sect.4 is
devoted to the inclusion of momentum operators and sect.5 deals with
the problem of physical invariants. The last section 6 contains our
conclusions.

\section{The model}

We consider a flat four-dimensional space time with operators $X^0, X^1, X^2,
X^3$ obeying the commutation relations
\bea [X^\l,X^\m]= i\Q^{\l\m},\qquad \Q^{\l\m}=-\Q^{\m\l}
\eea
where the $\Q^{\l\m}$ are considered to be "true" (real) numbers. Herewith we
mean that they are considered to be proportional to the unit operator $E$ of a
Lie algebra. Hence this operator has to commute with all other operators of
the algebra considered. It is this condition which restricts the model to a
great extent, what is viewed here as an advantage taking into account the huge
variety of models possible.

Therefore, besides the ordinary condition
\be [\Q^{\m\n},X^\l]=0 \ee
we demand for the commutator 
\be [\Q^{\m\n},M^{\r\s}]=0 , \ee
where the $M^{\r\s}$ (with $M^{\r\s}=-M^{\s\r}$) are the generators of the
Lorentz or $SO(3,1)$ group obeying the standard commutation relations among
each other. It might seem more natural instead of eq. (2.3) to postulate the
commutator to be given by a tensor representation of $SO(3,1)$ . Imposing
further Lie algebra conditions those models are consistent with commuting
momentum operators and allow the definition of mass and spin in the usual
manner. They are realized in standard noncommutative field theory (s. e.g.
[20,21], remind also some recently raised criticism baised on different
schemes of renormalization of a non-local field theory in [26]). Nevertheless this paper deals with models based on eq. (2.3) and is devoted
to studying the consequences of those conditions.

The set of physical operators which is postulated to form a Lie algebra $L$
contains 11 operators, besides the six Lorentz generators and the four space
time operators we also must keep the unit operator $E$ . So our algebra $L$
has a non-vanishing centre and is therefore nilpotent and hence solvable.
Due to a standard theorem in Lie algebra theory [22] $L$ can be decomposed into a
semi-direct sum
\be L=I_L\oplus)S \ee
where $I_L$ is the largest solvable ideal and $S$ is semi-simple. The part of
$S$ is obviously played here by the algebra $so(3,1)$ and therefore $I_L$ is
given by the set of operators $\{E,X^\m\}$ being nilpotent and hence solvable.

To fulfill the property of an ideal the commutators of $M$ and $X$ must
belong to $I_L$ :
\be [M^{\m\n},X^i]=i{f^{\m\n i}_j}X^j \ee

We have used Greek and Latin indices to manifest their different character in
further discussion. (The rather unphysical inclusion of $E$ on the r.h.s. of
eq. (2.5) has been disregarded.) We understand the Latin indices as operating in
a 4x4 matrix space while the Greek indices label the six different matrices
$F^A, A=1...6$ .
Then the semi-direct sum (2.4) forces those matrices (to be precise with a
factor $(-1)$) to form a representation of $so(3,1)$
\be [F^A,F^B]=-C^{AB}_CF^C \ee
where $C^{AB}_C$ are the structure constants of the algebra $so(3,1)$.
The last equation can be understood as the Jacobi condition for two $M$ and
one $X$ operators. The second condition derived from the semi-direct sum is
equivalent to the Jacobi identity for two $X$ and one $M$ operators. Taking
into account eq. (2.5) it can be written as a true matrix condition 
\be F\Q=-\Q{F^T} \ee
For clearness we add that the indices of $\Q$ should to be understood as Latin
ones.

Now we look for solutions of eqs. (2.6) and (2.7) with unknown matrices $F$ and
$\Q$ . Exploiting two times the symplectic algebra $sp(2)$ we have found the
solution
\be {\Q^{mn}}  = \left(\begin{array}{cccc}
0 & 0&A& B\\
0&0&B&-A\\
-A&-B&0&0\\
-B&A&0&0\end{array}       \right)
\ee
or explicitly
\bea \Q^{02}& =&-\Q^{13}= A ,\qquad \Q^{03}=\Q^{12}=B \\
     \Q^{01}&=&\Q^{23}= 0 \nn
\eea 
$A$ and $B$ are arbitrary real constants not fixed within our model.
The rank of $\Q$ is four as long as they do not vanish both. The
electric and magnetic components of $\Q$ are of equal value. 
The Lie algebra $F$ is composed of all matrices of the form
\be F= \left( \begin{array}{cccc}
J&-C&-D+K&-E-G\\
C&J&E+G&-D+K\\
D+K&-E+G&-J&C\\
E-G&D+K&-C&-J\end{array} \right)
\ee
where $C,D,E,J,G,K$ are arbitrary numbers.

The use of $sp(2)$ is due to the fact, that there is no non-trivial
solution, if $so(3)$ is used instead of $so(3,1)$ . The same happens if one
tries to substitute the last condition of eq. (2.9) by introducing a further
constant. 
The choice of the six matrices $F^A$ is not unique, we have taken the
$M$-operators to be antisymmetric and the $N$-operators to be symmetric. There
is still an obvious  freedom for a three-dimensional rotation.
Now we can give the explicit view of the $F$-matrices describing the
transformation of space time operators in our model. For shortness we write
$M$ and $N$ for the generators of the $F$-algebra but one has to keep in mind
that the "true" representation of $so(3,1)$ is given by $-M$ and $-N$ in our
notation.
\bea M^1= 1/2 \left( \begin{array}{cccc}
0&-1&0&0\\
1&0&0&0\\
0&0&0&1\\
0&0&-1&0\end{array} \right), \nn \qquad
M^2=1/2 \left( \begin{array}{cccc}
0&0&-1&0\\
0&0&0&-1\\
1&0&0&0\\
0&1&0&0 \end{array} \right) \nn \\
M^3=1/2 \left( \begin{array}{cccc}
0&0&0&-1\\
0&0&1&0\\
0&-1&0&0\\
1&0&0&0 \end{array} \right) ,\qquad
N^1=1/2 \left( \begin{array}{cccc}
-1&0&0&0\\
0&-1&0&0\\
0&0&1&0\\
0&0&0&1 \end{array} \right) \\
N^2=1/2 \left( \begin{array}{cccc}
0&0&0&1\\
0&0&-1&0\\
0&-1&0&0\\
1&0&0&0 \end{array} \right), \nn \qquad
N^3=1/2 \left( \begin{array}{cccc}
0&0&-1&0\\
0&0&0&-1\\
-1&0&0&0\\
0&-1&0&0 \end{array} \right) \nn
\eea
Those equations are to be read (s. eq.(2.5)) like
\bea
[M^1,X^0] = {-i \over 2} X^1, \qquad    
[M^1,X^2]= {i \over 2}X^3, \eea
\bea
[M^1,X^1] = {i \over 2} X^0, \qquad
[M^1,X^3]={-i \over 2} X^2 \nn
\eea
and analogously for $M^2,M^3,N^1,N^2,N^3$ .The first question is of course what
kind of representation of $so(3,1)$ is given by $F$ . In ordinary commutative
theory the space time operators are transformed by the usual vector
representation of $so(3,1)$ that is in the $D^{(i,j)}$ notation by the
representation $D^{(1/2,1/2)}$ . Our representation is reducible which can be
seen quickly by calculating the Lorentz invariant $\mathbf{\vec M \vec N}$ which is here
proportional to the matrix 
$ \left(\begin{array}{cccc}
0&-1&0&0\\
1&0&0&0\\
0&0&0&-1\\
0&0&1&0 \end{array} \right) $
excluding irreducibility.

Applying the non-singular transformation matrix 
\bea T={1 \over 2} \left(\begin{array}{cccc}
1&-i&0&0\\
0&0&1&-i\\
1&i&0&0\\
0&0&1&i \end{array} \right), \qquad
T^{-1}= \left(\begin{array}{cccc}
1&0&1&0\\
i&0&-i&0\\
0&1&0&1\\
0&i&0&-i \end{array} \right) \eea
the matrices (2.11) can be brought to the form $M'=TMT^{-1}$ 
\bea M'^1={i \over 2}\left(\begin{array}{cc}
-\s_z&0\\
0&\s_z \end{array} \right),
M'^2={i \over 2}\left(\begin{array}{cc}
-\s_y&0\\
0&-\s_y \end{array} \right),
M'^3= {i\over 2} \left(\begin{array}{cc}
-\s_x&0\\
0&\s_x \end{array} \right) \eea
\bea N'^1=-{1 \over 2}\left(\begin{array}{cc}
\s_z&0\\
0&\s_z \end{array} \right),
N'^2=-{1 \over 2} \left(\begin{array}{cc}
\s_y&0\\
0&-\s_y \end{array} \right),
N'^3=-{1 \over 2} \left(\begin{array}{cc}
\s_x&0\\
0&\s_x \end{array} \right) \nn \eea
which clearly shows the decomposition into $D^{(1/2,0)}\oplus D^{(0,1/2)}$,
the reducible representation used in Dirac's equation.
Nevertheless we mention that in our case both subspaces obtain the same
orientation only after an additional rotation around the $y$-axis in generator
space by the angle of $\p$ .
At the end of this section we comment about the relation of our model to other
ones. 

At first, exploiting another representation as usual for the transformation of
space time operators means to depart from viewing noncommutativity as a
deformation of commutative theory, i.e. for $\Q\rightarrow 0$ we do not obtain the
standard theory because there is no smooth way from an irreducible
representation to a reducible one (in general to any other non-equivalent).
We deeply believe that if live is noncommutative, there is no analyticity to
be expected in $\Q$ . This is quite analogous to perturbation theory in
ordinary quantum mechanics. We expect the picture to resemble some kind of
phase transition when one passes from commutativity to noncommutativity.

The second difference consists in the fact, that $\Q$ cannot be changed by
Lorentz transformations (s. eq. (2.3)) and therefore plays the role of a
constant external field, which of course breaks Lorentz invariance.
From eq. (2.8) it follows, that the $x$-dimension is treated differently from
$y$ and $z$, which are handled on the same footing. In a further study the
latter ones could be compactified. This leads to the idea to consider our
model as a toy model for higher compactified dimensions. Here the important
question arises wether noncommutativity might be connected with those extra
dimensions only.

\section{Explicit structure of our noncommutative Lie algebra}

For any finite-dimensional Lie algebra there exists a true matrix
representation. It is interesting to look at it for our five-dimensional
nilpotent algebra $I_L$ .
It can be easily seen that the five 4x4 matrices 
\bea X^0= \left(\begin{array}{cccc}
0&0&0&0\\
-B&0&0&0\\
0&0&0&0\\
0&0&A&0 \end{array} \right), \qquad
X^1= \left(\begin{array}{cccc}
0&0&0&0\\
A&0&0&0\\
0&0&0&0\\
0&0&B&0 \end{array} \right) \nn  \eea

\bea X^2= \left(\begin{array}{cccc}
0&0&0&0\\
0&0&0&0\\
1&0&0&0\\
0&0&0&0 \end{array} \right), \qquad
X^3= \left(\begin{array}{cccc}
0&0&0&0\\
0&0&0&0\\
0&0&0&0\\
0&1&0&0 \end{array}\right) \eea

\bea E= \left(\begin{array}{cccc}
0&0&0&0\\
0&0&0&0\\
0&0&0&0\\
1&0&0&0 \end{array}\right) \nn \eea 

fulfill the commutation relations

\bea [X^0,X^2]=A\cdot E, \qquad [X^0,X^3]=B\cdot E,\qquad 
[X^1,X^2]=B\cdot E, \eea \\
\bea [X^1,X^3]=-A\cdot E,\qquad [X^i,E]=0 \nn \eea

One can get rid of the constants $A$ and  $B$ after introducing so-called
renormalized light-cone coordinates $\tilde X^0$ and $\tilde X^1$ by
\bea \tilde X^0= \frac{-BX^0+AX^1}{A^2+B^2}, \qquad
\tilde X^1= \frac{AX^0+BX^1}{A^2+B^2} \eea
resulting in the easier commutation relations
\bea [\tilde X^0, X^3]=-1, \qquad [\tilde X^1, X^2]=1 \eea
All other commutators vanish. In the matrix representation $\tilde X^0$ and
$\tilde X^1$ take the form
\bea \tilde X^0= \left(\begin{array}{cccc}
0&0&0&0\\
1&0&0&0\\
0&0&0&0\\
0&0&0&0 \end{array}\right) \qquad
\tilde X^1=\left(\begin{array}{cccc}
0&0&0&0\\
0&0&0&0\\
0&0&0&0\\
0&0&1&0 \end{array}\right) \eea

As a next step we shall integrate the algebra $I_L$ and construct the Lie group 
(to be exact the universal covering Lie group) of the Lie algebra $I_L$ . We
call this five parameter group $ \tilde X^4$ which is the noncommutative
generalization of the four-dimensional space time coordinate group  . One way of
doing that is the following. Let $g\in  \tilde X^4$ be a function of the parameters
$y_0,\vec y$ and a phase $\f$ and 
\be g(y_0,\vec y,\f)=\exp (iX^0y_0+iX^1y_1+iX^2y_2+iX^3y_3+iE\f) \ee 

Using Baker-Campbell-Hausdorff's formula for operators whose commutators
commute with the original ones the composition law and the inverse element can
be defined in the way:
\bea g(y,\f)g(z,\s)&=&g(y+z,\f+\s+\frac{1}{2} iA[y_0z_2-y_2z_0-y_1z_3+y_3z_1]+\frac{1}{2}
iB[y_0z_3-y_3z_0+y_1z_2-y_2z_1]) \nn \\
g^{-1}&=&g(-y,-\f) \eea

Now one may ask how to construct finite dimensional irreducible
representations of the Lie group $\tilde X^4 \times )SO(3,1)$ . A standard
theorem [22] tells us that we have to look for all characters $\c$ of $\tilde X^4$
fulfilling the property
\be \c (g)=\c ({g_s}^{-1}gg_s) \ee
where \bea g_s\in SO(3,1) \nn \eea 
Eq. (3.8) is fulfilled only by characters depending on $\f$ solely. But the
composition law eq. (3.7) does not allow such characters to exist. Hence all
finite dimensional irreducible representations of our semi-direct product are
(like in commutative case) equivalent to the finite dimensional irreducible
representations of $SO(3,1)$ . We shall return to this important point after
the inclusion of momenta in the next section.

\section{The inclusion of momentum operators}

Now it is straightforward to include momentum operators in our
noncommutative Lie algebra. To do that in a consistent way one has to
care for all possible Jacobi conditions to be fulfilled by the
commutators invented. At first we suppose the commutators of $M$ and
$P$ operators to be given by a representation of $so(3,1)$, that is
\be [M^{\m\n},P^i]=ih_j^{\m\n i}P^j\ee
where the matrices $H^A$ have to obey the condition of eq. (2.6)
equivalent to ($MMP$) Jacobi identity. 
Next we wish to preserve the canonical quantization commutation
relations
\be [X^i,P^j]=-ig^{ij} \ee
where $g^{ij}$ is the Minkowski metric tensor. Now the ($MPX$) Jacobi
condition leads to the matrix constraint
\be H=-gF^Tg \ee
which determines $H$ fully. 

It follows in our model that momentum operator transformation is given
by an equivalent (but not identical) to coordinate operator
transformation representation of $so(3,1)$ .
From eq. (2.7) we derive 
\be FA=AH \ee with 
    $A=\Q g$        .(Remember that $A$ has a
non-vanishing determinant.) We remind the reader that in commutative
case for representation
$D^{(1/2,1/2)}$
$H$ is simply identical to $F$ . 

It is now consistent with our approach to assume
\be [P^\m ,P^\l ]=i \tilde \Q^{\m\n} \ee
and
\be [\tilde \Q^{\m\n},M^{\r\s}]=0 \ee
The concrete form of $\tilde \Q$ is up to now not specified, but the
($PPM$) Jacobi condition yields
\be H\tilde \Q=-\tilde \Q H^T \ee
Introducing $\tilde \Q'$ by $\tilde  \Q'=g\tilde \Q g$ eq. (4.7) is
equivalent to \be F^T\tilde \Q'=-\tilde \Q'F \ee
which can be easily solved for given $F$ from eq. (2.10). 
The result is
\be \tilde \Q = \left(\begin{array}{cccc}
0&0&C&D\\
0&0&-D&C\\
-C&D&0&0\\
-D&-C&0&0 \end{array}\right) \ee
where $C$ and $D$ are arbitrary (real) constants. The ($XXP$), ($PPX$)
and ($PPP$) Jacobi conditions are fulfilled if we put
\bea [\Q^{\m\n},P^i]=[\tilde \Q^{\m\n},X^i]=[\tilde \Q^{\m\n},P^i]=0
\eea
We stress the fact that our model is consistent even if $C$ and $D$
vanish both. In that case noncommutativity is connected with
coordinate space only and not with momentum space. This point reminds
strongly what happens in quantum field theory of noncommutative spaces
even though our model is different from theirs [23,24]. The consequences of
non-vanishing $C$ and $D$ will be analyzed in the next section.

\section{Invariants and the mass problem}

In this section we study the noncommutative generalization of the
Poincare group \\
$ \tilde T^4 \times )SO(3,1) $ where $\tilde T^4$  is the
generalization of the commutative four dimensional translation group $
T^4 $. Its Lie algebra is generated by the operators $P^0,P^1,P^2,P^3
$ and $E$ .
That algebra can be integrated in the same way as in section 3 .

Hence the Lie algebra of our generalized Poincare group is nilpotent
and therefore solvable. The usual construction of invariants for the
Lie group of our algebra does not apply because the Cartan-Weyl tensor
is singular. We expect the existence of two independent central
operators for our Lie algebra but in the literature [22,25] there was even no
theorem making statements about the number of invariants in case of
solvable Lie algebras. Therefore we shall construct below the bilinear
central operator by hand.

We consider the operator
\be \tilde I=i_{mn}P^mP^n \ee
with numbers $ i_{mn}$ forming the matrix $I$ . From eq. (4.5) we
calculate 
\be [\tilde I,P^l]=P^m(i_{mn}\tilde \Q^{nl}+ i^T_{mn}\tilde \Q^{nl}) \ee
In the same way after eq. (4.1) we find 
\be [\tilde I,M^{\m\l}]=-P^m(h^{T\m\l n}_m i_{nl} + i_{mn}h^{\m\l
n}_l)P^l \ee
which leads to the matrix condition 
\be H^TI=-IH \ee
This condition has to be fulfilled by all six matrices $H$ .
To solve this condition we apply a non-singular transformation matrix
$ \tilde T=gT^{-1T}g $ in the way 
\be H'=\tilde T H\tilde T^{-1} \ee
and therefore
\be I'= \tilde T^{-1T} I \tilde T^{-1} \ee 
We shall obtain $H'$ which are the six matrices of eqs. (2.14) with
several signs reversed. Nevertheless it is evident that any $I'$
fulfilling eq. (5.4) is given by 
\be I'= \left(\begin{array}{cc}
\s_y&0\\
0&\l \s_y \end{array}\right) \ee 
which after retransformation leads to 
\be I=\l_1 \left(\begin{array}{cc}
0&1\\
-1&0 \end{array}\right) + \l_2 \left(\begin{array}{cc}
0&\s_y\\
\s_y&0 \end{array}\right) \ee
with all $\l$ being arbitrary constants. (This $I$ is antisymmetric
and therefore obeys eq. (5.2).)
But now we have 
\be \tilde I=2i(\l_1C+i\l_2D) \ee
That means $ \tilde I \sim  E $, which may have been expected from the
very beginning. We have done the calculation in an explicit way to
show that here is no bilinear central operator besides $E$ . This is
one of the intrinsic problems of constructing central operators for a
nilpotent Lie algebra which by definition contains a non-trivial
centre. The only way to interprete the above result in a physical
manner is, that the mass operator is a constant and therefore all
particles have the same mass. This consequence, which at the first
moment seems to rule out the model, must be viewed in context with the
arguments of sect. 2 about a necessary phase transition. Then it is
really very unlikely that masses and other particle parameters
resemble each other on different sides of the transition point. (For
vanishing $C$ and $D$ the masses also vanish.) It is not surprising
that the square of the Pauli-Lubanski vector $\e_{lmnk}M^{mn}P^k$ is
not central in our Lie algebra. By direct calculation we have
established that the commutator with $P^s$  does not vanish (It is a
non-zero combination of $\tilde \Q$, $M$ and $P$ .) while the one with
$M^{rs}$ does. 

To construct infinit dimensional irreducible representations of our
generalized Poincare group one has to adjust the standard procedure
(for Abelian $T^4$) via orbits to the noncommutative case which seems
to be an interesting task for the future.

\section{Conclusions}
We have presented a Lie-algebra model for noncommutative coordinate
and momentum space which contains four unrelated parameters not to be
determined within the model. After redefinition the number of
parameters can be reduced to two, namely $\sqrt{A^2+B^2}$ and
$\sqrt{C^2+D^2}$ , the former setting the scale for $\Q$ and the latter
for the mass. 

While Lorentz covariance is broken by $\Q$, $\Q$ itself is considered
as a Lorentz scalar.

The representation under which coordinate and momentum operators are
transformed can be no longer $D^{(1/2,1/2)}$ , the usual vector
representation. It turns out that the easiest possibility is the
reducible spinor representation $D^{(1/2,0)}\oplus   D^{(0,1/2)}$ , well known from
Dirac's equation. That forces us to adopt the occurence of a phase
transition between noncommutative and commutative world instead of the
usual conception of a smooth deformation. This point has to be worked
out in further research. The phase transition also has to generate the
mass spectrum because in our approach the only suitable mass operator
is a constant. The main open question is the construction of a second
invariant which is thought to replace ordinary spin. This problem is
under consideration now.

\section{Acknowledgements}
This work has been supported by DFG . \\
The author thanks J. Wess, D. L\"ust, H. Dorn and M. Karowski for
helpful discussions and J. Lukierski for critical advice.

\end{document}